\documentclass[apjfonts]{emulateapj}

\slugcomment{Submitted to The Astrophysical Journal}

\shorttitle{Bright X-Ray Sources in Elliptical Galaxies}
\shortauthors{Ivanova \& Kalogera}

\begin{document}

\title{The Brightest Point X-Ray Sources in Elliptical Galaxies \\ and the Mass Spectrum of Accreting Black Holes} 

\author{N.\ Ivanova\altaffilmark{1},  V.\ Kalogera\altaffilmark{1}} 

\affil{$^{1}$ Northwestern University, Dept of Physics and Astronomy,
2145 Sheridan Rd, Evanston, IL 60208, USA.}

\email{nata and vicky@northwestern.edu}

\begin{abstract}
We propose that  the shape of the upper-end  X-ray luminosity function
(XLF) observed  in elliptical galaxies for  point sources carries
valuable information  about the  black-hole (BH) mass  spectrum among
old X-ray transients formed in  the galaxies. Here we present the
line of  arguments and analysis  that support this connection  and the
methodology for deriving the BH  mass spectrum slope from the observed
XLF slope. We show that  this underlying BH mass spectrum is modified
by a weighting factor that is  related to the transient duty cycle and
it generally depends on the host-galaxy age, the BH mass and XRB donor
type (main-sequence,  red-giant, or white-dwarf  donors). We find
that the observed XLF is dominated by transient BH systems in outburst
(a prediction possibly testable  by future observations), but that the
assumption of a  constant duty cycle for all  systems leads to results
inconsistent with current observations. We also find that the derived
BH mass slope depends on the  strength of angular momentum loss due to
magnetic braking for main-sequence  donors. More specifically, we find
that, for ``standard'' magnetic braking, BH XRBs with red-giant donors
dominate the upper-end XLF;  for weaker magnetic braking prescriptions
main-sequence donors  are found  to be dominant.  The methodology
presented here can  be used in the future as  our understanding of the
transient  duty  and  its   dependence  on  binary  and  mass-transfer
properties improves. Under certain  assumptions for this dependence we
derive a differential BH mass spectrum slope of $\simeq 2.5$; an upper
BH mass  cut-off at $\simeq  20$\,M$_{\odot}$ is needed  to understand
the very brightest of the BH XRBs in elliptical galaxies. We also show
that our  quantitative results are robust  against expected variations
by factors of a few of the outburst peak X-ray luminosities. We expect
that our analysis will  eventually help to constrain binary population
synthesis models and the adopted relations between black holes and the
masses of their progenitors.
\end{abstract}


\section{INTRODUCTION}

{\em Chandra} has  revolutionized the study of point  X-ray sources in
the nearby Universe. The majority of these are interpreted to be X-ray
binaries  \citep[XRBs;   for  a   general  review  on   {\em  Chandra}
performance  see  Weisskopf  et  al.\ 2003,  for  extragalactic  X-ray
binaries   see,   e.g.,][]{2004ApJ...611..846K,  2004ApJ...613..279J}.
Elliptical galaxies out to the Virgo cluster have now been studied and
have  surprised us  with the  large number  (typically $\sim  100$ per
galaxy) of  detectable point X-ray sources down  to X-ray luminosities
of    about    {$10^{37}$\,erg\,s$^{-1}$.    Two    main    population
characteristics have attracted considerable  attention so far: (i) the
X-ray luminosity function (hereafter XLF)  that may or may not exhibit
a break at about $4-5\times10^{38}$\,erg\,s$^{-1}$~\citep[for a recent
update see][]{2004ApJ...611..846K}; (ii)  the high fraction of sources
coincident with identified globular clusters (GCs) in ellipticals.

The shape of the XLF has been debated since the first observations of
ellipticals were reported.  \cite{2000ApJ...544L.101S} identified a
shape that required two power laws with a ``break'' or a ``knee'' at
$\simeq\,3.2\times\,10^{38}$\,ergs\,s$^{-1}$.
\cite{2003ApJ...586..826K} argued that the break may result from 
biases affecting the detection threshold of
the data.  In the following few years longer exposures became possible
and more and more ellipticals were added in the observed sample with
low enough sensitivity \citep[see, e.g.][]{2004MNRAS.349..146G}.  
The current situation is probably best
summarized in \cite{2004ApJ...611..846K}. They analyzed a large
sample of elliptical galaxies with varying sizes of point source
samples, and they concluded that: although XLFs of individual galaxies
do not require a broken-power-law fit, the combined sample of sources
from all the galaxies considered shows a statistically significant
requirement for two power laws and a break at $5\pm 1.6\times
10^{38}$\ ergs\ s$^{-1}$.  They found the best-fit slope of the lower
end of the differential XLF to be $\alpha_{\rm d}=1.8\pm0.2$ and the best-fit slope of the
upper end to be $\alpha_{\rm d}=2.8\pm0.6$.  It is important to note that the
break location is consistent  with the Eddington luminosity for a
$1.9\pm0.6$\,M$_{\odot}$ neutron star (NS) accreting 
helium-rich material (for hydrogen rich donor this value is 
as large as $3.2 \pm 1 M_\odot$).  In what follows we
consider the results of the \cite{2004ApJ...611..846K} study as
representing our current observational understanding of the XLF in
ellipticals.  We address the question of the interpretation of this
understanding and what it implies about the properties of the sources
contributing to the observed XLFs. 

Large fractions (20\% -- 70\%) of the point sources in ellipticals have been reported 
to be associated with globular clusters \citep[see, e.g.,][and references therein]{2003ApJ...595..743S}.
These high fractions have led to the suggestion that {\em
all} point X-ray sources seen currently in ellipticals have been
formed through stellar interactions and that sources that are not
associated with GCs have originated in GCs and have either been
ejected or the parent GCs have been destroyed by the galaxian tidal
forces.  As much interesting as this suggestion is, it raises the
question: why would the field stellar population of ellipticals not
lead to XRB formation as it has occurred in the Milky Way, for
example?  One could speculate that the field population is just too
old and the XRBs that were formed at some point have completed their
X-ray emitting life. However, such a speculation is inconsistent with
the expectation that XRBs with low mass donors can live for several
Gyrs as the donors lose mass and the binaries enter a transient
phase. Such systems would become detectable as bright X-ray sources
during the disk outbursts \citep{2002ApJ...571L.103P}.  Moreover, it
has recently been pointed out that the high rate of source coincidence
with GCs actually appears consistent with some of the sources having
been formed in the field \citep{2005ApJ...621L..25J}, although the
result may sensitively depend on the definition of GC concentration in
ellipticals. Most importantly for the present study, of the bright sources 
above the XLF break only a very small fraction is associated with GCs 
\citep[e.g., only one source in the Virgo cluster sample as reported by][]{2004ApJ...613..279J}.

In a recent {\em Letter} \citet[][hereafter DB]{2004ApJ...607L.119B} 
have suggested an explanation that couples the two population
characteristics: the XLF shape and the source coincidence with GCs. 
They identify the point X-ray sources as ultra-compact binaries (UCBs) 
that form predominantly in GCs.  
They are neutron stars accreting from low-mass He or C/O white dwarfs 
and they contribute to the ellipticals XLF early in their lifetime 
when they are still bright. 
 Their association with GCs is important in replenishing the population 
through tidal interactions and allowing a significant number of 
sources in this bright phase, even though there is no ongoing star 
formation in ellipticals nor in GCs. 
DB estimate the rate of ultracompact binary formation to be consistent with 
the number sources observed and conclude that the model XLF slope is 
in agreement with the slope {\em below the break} as derived by
\cite{2004ApJ...611..846K}.  The association of the XLF break with the
NS Eddington limit for He-rich accretion is also consistent with this
interpretation.  However, the {\em upper end} of the XLF (at
luminosities in excess of the break location) are not easy to
interpret.  Deloye \& Bildsten suggest that some of the NS
ultra-compact sources can reach super-Eddington luminosities. However
luminosities in excess of $10^{39}$\,erg\,s$^{-1}$ are very difficult to 
explain with NS accretors. 
Therefore the origin of the upper-end slope is not naturally connected to NS UCBs formed in GCs. 

In this paper we address the question of the {\em upper-end} XLF slope and its origin.  
We consider the previously made suggestion
\citep{2000ApJ...544L.101S} that the XLF above the break at $5\times
10^{38}$ erg s$^{-1}$ is populated by XRBs with black hole (BH) accretors. 
Given that GCs are not expected to harbor a significant number of BH-XRBs 
\citep[see][and references therein]{2004ApJ...601L.171K}, 
we suggest that the vast majority of these BH-XRBs are part
of the galactic-field stellar population in ellipticals. 
As we will show most of donors in these binaries are of low-enough mass that the 
XRBs are expected to be transient and therefore they populate the XLF only during disk 
outbursts when they typically emit at the Eddington luminosity for their BH mass. 
We further suggest that the slope of the upper XLF is a footprint of the BH mass 
spectrum in the BH XRBs under consideration. 
We present analytical derivations that demonstrate this link and we develop a method that allows us to infer 
the underlying BH mass spectrum consistent with the current upper-end XLF slope~\citep{2004ApJ...611..846K}.  
We also show that given the current observations it is possible to 
constrain the strength of magnetic braking acting in these XRBs, 
the type of BH donors, as well as the transient duty cycle to some extent. 
We also examine the quantitative robustness of our results against variations of some basic assumptions.  
This analysis is presented in \S\,2 and 3. 
We conclude with a discussion of our results and possible connections to population synthesis calculations (\S\,4).

\section{Black Hole X-Ray Binaries in Ellipticals}

We consider XRBs that could possibly populate the part of the observed
XLF above the reported break at $4-6\times\,10^{38}$ erg s$^{-1}$, and
therefore we focus on BH accretors (masses in excess of
$2-3$\,M$_{\odot}$).  Given the current estimates for the ages of
stellar populations in ellipticals \citep[in their majority 8 to 12
Gyr, although some estimates are slightly shorter than
5\,Gyr; see][]{2001MNRAS.326.1141R, 2005ApJ...622..235T}, we expect that
donor masses are lower than $\simeq 1-1.5$\,M$_{\odot}$.  Given these
mass ratios and the properties of similar observed systems in the
Milky Way (i.e., soft X-ray transients), these BH XRBs are expected to
be transient X-ray sources (see McClintock \& Remillard 2005 for a review of BH X-ray binaries in
the Milky Way), where mass transfer is driven by the
Roche-lobe filling donor.  Given the above upper limit on the donor
mass for ellipticals, we expect that there will be three different
types of low-mass donors: (i) Main Sequence (MS) stars, (ii) Evolved
or Red Giant Branch (RG) stars, and (iii) White Dwarf (WD)
donors. Each of these sub-populations of BH XRBs will have different
typical mass-transfer rates and binary property distributions, and
therefore we examine them separately in our analysis that follows. 

\subsection{Transient X-Ray Sources} 

We  adopt  the  current  understanding  for the  origin  of  transient
behavior in XRBs \citep[for a recent review, see][]{BH_book_ch13}.  To
identify  transient systems in  our modeling  we consider  the typical
mass-transfer  (MT)  rate  associated  with  each type  of  XRB  donor
($\dot{M}_{i}$) and compare it to  the critical MT rates for transient
behavior  ($\dot{M}_{\rm  crit}$):   if  $\dot{M}_{i}  <  \dot{M}_{\rm
crit}$, then the  accretion disk is expected to  be thermally unstable
and  the binary  system is  assumed to  be a  transient  X-ray source.
The value  of this critical MT  rate for the  disk instability is
not precisely known  and its functional dependence on  disk and binary
properties are  subject to  uncertainties associated with  our current
theoretical understanding  of the disk instability.  However, both the
qualitative  concept of  the  existence  of a  critical  rate for  the
instability to set in and its quantitative estimates by recent studies
appear to  be in  good agreement with  the behavior of  Galactic X-ray
transients.  Therefore,   we  adopt  the   current  understanding  and
quantitative estimates. More  specifically, for hydrogen-rich donors,
we    adopt    the    $\dot{M}_{\rm    crit}$   value    derived    by
\cite{1999MNRAS.303..139D}, and for helium or carbon-oxygen donors, we
adopt  the  value  derived  by \cite{2002ApJ...564L..81M}.   The
effects of  quantitative deviations  from the adopted  expressions are
discussed in what follows.

To account for the contribution of transient sources in the XLF among
any persistent sources, assumptions about the XRB luminosity at outburst 
and the transient duty cycle need to be made. 

When a XRB is identified as transient, we assume that during the disk
outburst the X-ray luminosity $L_{\rm X}$ is equal to the Eddington
luminosity $L_{\rm Edd}$ associated to the BH accretor:
 \begin{equation} 
 L_{\rm Edd} = \frac{ 4\pi c G M_{\rm BH}}{\kappa} = 
 5\times 10^{37}\,\frac{m_{\rm BH} }{ \kappa} \ \ {\rm ergs\,s^{ -1}}\ ,
 \end{equation}
 where $m_{\rm BH}$ is the accretor mass in $M_\odot$ and $\kappa$ is
the opacity of the accreting material in cm$^2$\,g$^{-1}$. 
We adopt electron scattering
opacities equal to 0.32 and 0.19 for hydrogen and helium or
carbon-oxygen rich material, respectively.  

We note that of the 15
confirmed transient BH XRBs in our Galaxy, 3 appear to reach possibly
super-Eddington luminosities~\citep{BH_book_ch4} at outburst 
(although distance estimate uncertainties cannot be ignored).  Two of
them, V4641~Sgr and 4U~1543-47, have early-type donors.  Such donors
are not present in elliptical galaxies with population ages of $\sim
5-10$ Gyr.  The third one, GRS~1915+105, has a low-mass giant donor and an orbital period of 33 days. 
However its X-ray luminosity at outburst just barely exceeds its $L_{\rm Edd}$, by 40\% only.  
Given distance uncertainties associated with such an estimate, 
we conclude that we can neglect the possibility of super-Eddington luminosities 
during outburst in our XLF modeling. On the other hand outburst peak luminosities 
cover a significant range at sub-Eddington values. During primary\footnote{We use the term 
``primary'' to distinguish from ``follow-up'' outbursts that are occasionally observed 
very soon after primary ones with peak luminosities orders of magnitude below the 
Eddington limit (Remillard \& McClintock 2005, private communication). 
Such small outbursts do not reflect an extremely short duty cycle and do not 
contribute to the high-end XLF of interest here.} outbursts peak $L_X$ values 
can be lower than $L_{\rm Edd}$ by factors of a few (McClintock \& Remillard 2005, private communication). 
As part of our analysis we examine the effect of such variations on the methodology 
and conclusions presented here (see \S\,3.1.1). From an observationally point 
of view it has been shown (Zezas et al. 2004 and Zezas 2005, private communication) 
that variability in X-ray fluxes (and hence luminosities) by factors of a few 
(typical among accreting sources and detected with {\em Chandra} observations 
at different epochs) do not alter the XLF slopes as measured for nearby galaxies 
within the current errors. Therefore observationally the reported XLF slopes 
appear to be robust. Consequently we can use them to learn about the underlying 
XRB population with considerable confidence.

At present there are no strong constraints on the duty
cycles either from observations or from theoretical considerations. 
Among known Galactic X-ray transients, typical duty cycles of a few\,
\% is favored for hydrogen donors~\citep{1996ARA&A..34..607T}.  
To our knowledge, there are no data on duty cycles for transients 
with a WD companion. In what follows we investigate how plausible duty cycle
assumptions affect the upper-end XLF shape.  In particular, we
consider two specific cases: one of constant duty cycle equal to
$\eta=0.01$; another of a variable (dependent on MT rates) duty cycle
equal to 
 \begin{equation}
\eta=0.1\,\left(\frac{\dot M_{\rm  d}}{\dot M_{\rm crit}}\right)^{\,\delta}\,  
\label{eq:eta}
 \end{equation}
 where  $\delta=1$  is assumed.  The  first  of  these two  cases
 corresponds to the standard assumption of a constant duty cycle often
 made in the literature. The second case is motivated primarily by our
 plan to  examine how  one example form  of a MT-dependent  duty cycle
 affects our  analysis and results. Admittedly the  specific choice of
 the  dependence on  $\dot M_{\rm  crit}$ shown  above is  not solidly
 motivated,   given   all    the   uncertainties   of   the   outburst
 mechanism. However  it implies a  correlation of the duty  cycle with
 how  strong a  transient the  system is:  the further  away  from the
 critical MT rate, the smaller the  duty cycle. We stress that in most
 of our  analysis we  adopt this form  with $\delta=1$ as  a plausible
 example and  throughout the  paper we contrast  the results  to those
 obtained with  a constant duty  cycle. Furthermore in  \S\,\ref{monte-carlo}
  we examine the  sensitivity of our
 results  for main-sequence  donors on  the choice  of  the duty-cycle
 dependence   on  MT   extensively:   we  adopt   $\delta   >  1$   in
 eq.(\ref{eq:eta}) and  we also introduce  yet one other example  of a
 MT-dependent duty cycle:
 \begin{equation}
\eta=0.1\,\left(\frac{\dot M_{\rm  d}}{\dot M_{\rm EDD}}\right)^{\,\delta}\, 
\label{eq:eta2}
 \end{equation}
for which we  examine various values of $\delta$.  Once again there is
no   solid   theoretical  motivation   for   this  latter   functional
choice. However,  it provides  us with a  better understanding  of how
sensitive our  results are to  the details of the  possible duty-cycle
dependence on MT properties.

\subsection{Main Sequence Donors}
                                
Mass transfer  in BH  XRBs with  hydrogen-rich, low-mass MS  donors is
expected  to be  driven by  angular  momentum losses  due to  magnetic
braking (MB) and gravitational radiation (GR). In the case of
conservative   mass   transfer~\citep{1993ARA&A..31...93V}  the angular momentum loss rate are connected to the MT rate as follows:
 \begin{equation}
 \frac{\dot J_{\rm gr}}{J_{\rm orb}} + \frac{\dot J_{\rm mb}}{J_{\rm orb}}
= \frac{\dot M_{\rm d}}{M_{\rm d}} \left ( \frac{5}{6} + \frac{n}{2} - \frac{M_{\rm d}} {M_{\rm BH}}\right ),  
\label{jcons}
 \end{equation}
 where $n$ is the radius-mass exponent for the donor. For a low-mass MS star: 
 \begin{equation}
 r_{\rm d} \simeq   m_{\rm d}, 
\label{mr-rel}
 \end{equation}
 where  $r_{\rm d} = R_{\rm d}/ R_\odot$ and $m_{\rm d}=M_{\rm d}/M_\odot$  are the donor stellar radius and mass in solar units. 
Therefore $n\equiv\,d\ ln R_{\rm d} / d\ ln M_{\rm d}$ is equal to $\simeq1$ for MS donors. 

According to general relativity the rate of angular momentum loss due to GR is given by:  
 \begin{eqnarray}
\frac{\dot J_{\rm gr}}{J_{\rm orb}} =  - \frac{32 G^3}{5c^5}\,\frac{M_{\rm BH}M_{\rm d}(M_{\rm BH} + M_{\rm d})}{ A^{4}} 
\ \ \ \ \ \ \ \ \ \ \ \ \nonumber \\ 
\ \ \ \ \ \ \ \ \ \ \ \ =  - 2.6 \times 10^{-17}\,\frac{m_{\rm BH} m_{\rm d}(m_{\rm BH} + m_{\rm d})}{a^{4}} \,{\rm s}^{-1}, 
\label{jgr}
 \end{eqnarray}
 where $a$ is the orbital semi-major axis in units of solar radius. 
For a  mass ratio $q\equiv M_{d}/M_{\rm BH}<0.8$ \citep{1971ARA&A...9..183P} 
and using the mass-radius relation eq.~(\ref{mr-rel}): 
 \begin{equation}
 a =\frac{1}{0.46} m_{\rm d}^{2/3} (m_{\rm BH}+m_{\rm d})^{1/3}.
\label{rrl}
\end{equation}

We consider two derivations of the angular momentum loss rate due to magnetic braking: 
(i) the Skumanich law based on the empirical relation for slowly rotating stars 
adopted from  \citep{1983ApJ...275..713R} (RVJ), and 
(ii)  the revised law based on X-ray observations of faster rotating 
dwarfs adopted from \citep{2003ApJ...599..516I} (IT):
 \begin{eqnarray}
 & \dot J_{\rm mb}^{\rm RVJ} & =  -3.8\times 10^{-30} 
M_{\rm d} R_{\odot}^4  (R_{\rm d}/R_\odot)^2 \Omega^3 \ {\rm dyn\, cm} \, \\
 & \dot J_{\rm mb}^{\rm IT} & =  -6\times 10^{30}  (R_{\rm d}/R_\odot)^4 
\left ( \frac{\Omega^{1.3}  \Omega_{\rm x}^{1.7}}{\Omega_\odot^{3}}\right ) \ {\rm dyn\, cm}, 
 \end{eqnarray}
 where $\Omega$ [s$^{-1}$] is the stellar  angular velocity  which is equal to the
 binary orbital velocity assuming  the star is in full synchronization
 with the  binary orbit, $\Omega_\odot = 5\times  10^{-6}$ s$^{-1}$ is
 the      Sun's      angular      velocity,      and      $\Omega_{\rm
 x}=10\Omega_\odot$. Using Kepler's law the above are re-written as:
 \begin{equation}
 \frac{\dot J_{\rm mb}^{\rm RVJ}}{J_{\rm orb}} = -7.2\times 10^{-15} \,\frac{m_{\rm d}^2 (m_{\rm BH}+m_{\rm d})^2}{m_{\rm BH} a^5} \,{\rm s}^{-1} \ , 
\label{jvz}
 \end{equation}
 \begin{equation}
\ \ \ \  
\frac{\dot J_{\rm mb}^{\rm IT}}{J_{\rm orb}} = -2.7\times 10^{-17} \,\frac{m_{\rm d}^3 
(m_{\rm BH}+m_{\rm d})^{1.15}}{m_{\rm BH} a^{2.45}}\,{\rm s}^{-1} \ . 
\label{jit}
 \end{equation}
 In XRBs with BH masses $\ge 3 M_\odot$ and MS donor masses $\le 1.0 M_\odot$, it is  
$J_{\rm gr}\ll J_{\rm mb}^{\rm RVJ}$ and $J_{\rm gr} \ga J_{\rm mb}^{\rm IT}$. 
The lifetime of the MS-BH XRBs is much longer in the latter case.

As mentioned earlier in this study we adopt the derivation of the critical 
MT rate below which the accretion disk becomes unstable for irradiated disks presented by Dubus et al.\ 1999: 
 \begin{equation}
 \dot M_{\rm crit} = -1.5 \times 10^{15}  m_{\rm BH}^{-0.4} 
\left(\frac{R_{\rm disk}}{10^{10}{\rm cm}}\right)^{2.1}{\rm g\, s^{-1}} \ .
\label{mcrit_h}
 \end{equation}
Here $R_{\rm disk}$ is the  radius of the accretion disk. We note
that the exact value of this critical rate is subject to uncertainties
associated with our limited understanding of the disk instability, but
we  adopt the above  expression as  indicative of  the process  and we
continue with our analysis.

\begin{figure}
{\plotone{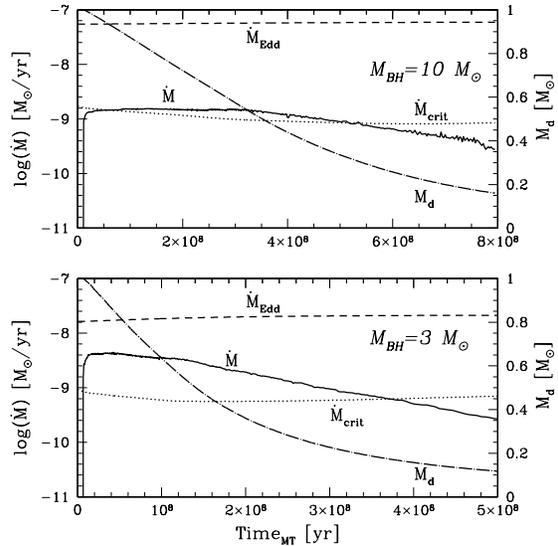}}
 \caption{Evolution of BH-MS binaries with the RVJ magnetic braking prescription, 
for BHs of 3 and 10 $M_\odot$; the initial MS companion mass is 1  $M_\odot$.
Shown are the MT rate $\dot M$ (solid line), the critical MT rate $\dot M_{\rm crit}$ 
(dotted line) and the Eddington MT rate $\dot M_{\rm Edd}$  (dashed line),  all rates are in $M_\odot$ per yr.
The dash-dotted line shows the donor mass evolution.}
\label{mt_rvj}
\end{figure}

From eq.(\ref{jcons}), (\ref{jvz}) and (\ref{mcrit_h}) it can be shown
numerically, that for the RVJ MB law and for low-mass donors, there is
a  BH mass  $M_{\rm PT}$  of  $\sim 5\,M_\odot$  that separates  BH-MS
systems  into persistent ($M_{\rm  BH} <  M_{\rm PT}$),  and transient
($M_{\rm  BH}   >  M_{\rm  PT}$).   From   more  detailed  binary
evolutionary  calculations using  the  stellar evolution  and MT  code
described in  \citep{2004ApJ...601.1058I}, we find  that this boundary
is about  10\,$M_\odot$ (see Fig.~\ref{mt_rvj} for  details).  For BHs
less massive than this critical  mass, the XRBs are persistent as long
as the  donor masses  are higher than  about 0.3\,$M_\odot$.  In these
persistent sources the MT rates driven  by the RVJ type of MB turn out
to be $0.01-0.25$  of the black holes's Eddington  rate.  As a result,
the   persistent  X-ray   luminosity   for  these   systems  is   $\la
10^{38}$\,erg\,s$^{-1}$,  i.e.,  below  the  bright $L_{X}$  range  we
consider  here.  Therefore  we  conclude  that  the  persistent  BH-MS
binaries driven by the RVJ  type of MB cannot contribute significantly
to the upper-end XLFs of ellipticals.

Next  we examine whether  the transient  phases associated  with BH-MS
binaries  and the  RVJ  MB law  are  important when  they reach  X-ray
luminosities  comparable to  the Eddington  limit. For  $M_{\rm  BH} >
10$\,$M_\odot$, the outburst luminosity is expected to be in excess of
$\simeq  1.5\times10^{39}$\,erg\,s$^{-1}$.  However, this  lower limit
is  comparable to  the highest  luminosity seen  currently in  XLFs of
ellipticals~\citep{2004ApJ...611..846K},  and therefore  these systems
cannot  contribute  significantly  to  the  observed  XLFs.  The  last
possibility  is outbursts  from  transient BH-MS  with  $M_{\rm BH}  <
10$\,$M_\odot$    and    donors     less    massive    than    $\simeq
0.3$\,M$_\odot$. Such  low mass donors are out  of thermal equilibrium
and   significantly  expanded  ($\simeq   3\times$)  compared   to  an
undisturbed MS star of the same mass.  In the case of the MT dependent
duty-cycle  $\eta$  is  about  a   few  \%.  We  note,  however,  that
applicability of  MB for  these stars is  very questionable, as  it is
generally accepted that MB does not operate in fully convective stars,
which  are  found  to  be  less  massive  than  $0.35$\,M$_\odot$  for
undisturbed     stars.    In    principle,     however,    ``eroded'',
out-of-thermal-equilibrium low-mass MS donors like the ones in XRBs do
not necessarily become  fully convective at the same  critical mass as
stars  with  no prior  MT  evolution. For  this  reason  we have  used
detailed MT calculations  (Fig.~\ref{mt_rvj}) with a stellar-evolution
code to examine this behavior  further. We find that in the donor-mass
range $0.15-0.3  M_\odot$ the radiative core is  extremely small, even
for these  ``eroded'' stars,  and therefore applying  angular momentum
loss due to  MB is not reasonable. Instead MB  activity and hence mass
transfer  is expected  to be  interrupted until  eventually  GR drives
Roche-lobe overflow much later.

Based on the  above line of arguments we  conclude that BH-MS binaries
evolving according  to the RVJ MB  law are not  expected to contribute
significantly to the high-end XLFs of ellipticals.
 
\begin{figure}
{\plotone{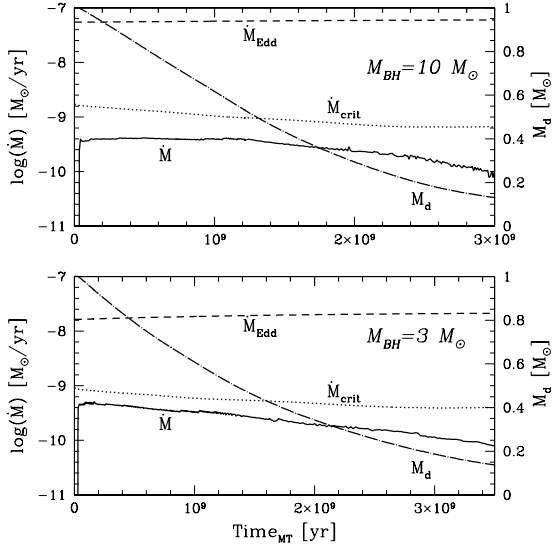}}
 \caption{Evolution of BH-MS binaries with the IT magnetic braking prescription, 
for BHs of 3 and 10 $M_\odot$; the initial MS companion mass is 1  $M_\odot$. 
Shown are the MT rate $\dot M$ (solid line), the critical MT rate $\dot M_{\rm crit}$ 
(dotted line) and the Eddington MT rate $\dot M_{\rm Edd}$  (dashed line),  all rates are in $M_\odot$ per yr.
The dash-dotted line shows the donor mass evolution.}
\label{mt_it}
\end{figure}

In the case of the IT MB prescription, BH-MS systems are transient for
all  BHs  masses $M_{\rm  BH}>  3 M_\odot$  and  for  all low-mass  MS
donors.  The reason  is that  the IT  MB is  weaker;  consequently the
donors are  mildly out  of thermal equilibrium  and the  mass transfer
rates are lower compared to the  RVJ MB case.  Using again detailed MT
evolutionary simulations we find that $\dot M /\dot M_{\rm crit}\simeq
0.25\pm0.15$ (see Fig.~\ref{mt_it}).  Therefore, in the case of the MT
dependent duty-cycle  $\eta$ is  again about a  few \%. In  both cases
(RVJ MB  and IT MB),  the value of  the duty cycle is  consistent with
observations for BHs of different masses, though the transiency occurs
at very different donor masses.

We  conclude that, regardless  of the  specific MB  law, it  is rather
unlikely   that  persistent  sources   with  a   BH  accretor   and  a
hydrogen-rich, low-mass MS donor  populate at any significant fraction
the upper-end  XLF of ellipticals; only  transient BH-MS sources
driven by the IT MB law can populate this X-ray luminosity range.

\subsection{Red Giant Donors}

For orbital periods more than about a day, MT occurs when the
low-mass donor is a subgiant or a giant.  The driving force is the
nuclear expansion of the donor, and a simple analytic prescription for
the MT is (Webbink, Rappaport, Savonije 1983; Ritter 1999; see also
King 2005):
 \begin{equation}
 \dot M{\rm rg} = - 3.4 \times 10^{15} a^{1.4} \frac{m_{\rm d}^{1.47}}
{(m_{\rm BH}+m_{\rm d})^{0.465}}\ {\rm g\, s^{-1}} \ .
\label{dmrg}
 \end{equation}
 It has been shown \citep[by][]{1997ApJ...484..844K,2000MNRAS.312L..39K}  that  such  wider XRBs
are transient, regardless of the BH mass  
(original derivations were based on a somewhat different 
expression for the critical MT rate for transient behavior, but still quite similar to eq.~[\ref{mcrit_h}].

\subsection{White Dwarf Donors}

A typical WD mass-radius relation is~\citep[see, e.g.,][]{1987ApJ...322..842R}:
 \begin{equation}
r_{\rm d} =  0.0128  m_{\rm d}^{-1/3}
 \end{equation}
  Using an approximation for the Roche Lobe radius \citep[from][]{1971ARA&A...9..183P} 
and assuming that the mass of WD is much smaller than a BH mass, we can show that
 \begin{equation}
a =  0.0278 \,   m_{\rm d}^{-1/3} \left ( \frac{m_{\rm d}}{m_{\rm BH}}\right )^{-1/3}
\label{a_wd}
 \end{equation}
We consider again conservative mass transfer (\ref{jcons}) but without any MB losses,
adopting  $n=-1/3$ and assuming that $m_{\rm d}\ll m_{\rm BH}$. Then
 \begin{equation}
\dot M_{\rm d} \simeq  
- \frac{32 G^3}{3.3c^5} \frac{M_{\rm BH}^2 M_{\rm d}^2 }{A^4} = 
- 7.9\times 10^{16} \frac{m_{\rm BH}^2 m_{\rm d}^2 }{a^4} \  {\rm g\, s^{-1}}
 \end{equation}
 We substitute here eq.~(\ref{a_wd}) and then have
 \begin{equation}
\dot m_{\rm d} = - 2 \times 10^{-3}\, m_{\rm d}^{4 \frac{2}{3}} m_{\rm BH}^{\frac{2}{3}}\ \, M_\odot {\rm yr^{-1}}
\label{dm_wd}
 \end{equation}

In what follows we adopt the critical MT rate for He-rich 
donors from \citet{2002ApJ...564L..81M}, 
but we note that the expression is subject to quantitative uncertainties associated 
with the current understanding of the disk instability:
 \begin{equation}
\dot M_{\rm crit} = -5.9 \times 10^{16}  m_{\rm BH}^{-0.87} 
\left(\frac{R_{\rm disk}}{10^{10}}\right)^{2.62}{\rm g\, s^{-1}}
 \end{equation}
 For a large mass ratio $q_{\rm BH} = m_{\rm BH}/m_{\rm d}$ 
the Roche lobe of the accretor is $r_{\rm RL} \simeq 0.7 a$ and
 \begin{equation}
r_{\rm disk} \simeq 2/3 r_{\rm RL}  = 0.013   m_{\rm d}^{-2/3} m_{\rm BH}^{1/3}
\label{rdisk}
 \end{equation}
 \begin{equation}
\dot m_{\rm crit} = -1.7 \times 10^{-12} m_{\rm d}^{-1.74} \ M_\odot {\rm yr^{-1}}
\label{mcrit_he}
 \end{equation}

BH-WD binaries will be transient if
 \begin{equation}
\frac{\dot m_{\rm d}}{\dot m_{\rm crit}} =  1.2\times 10^{9}  m_{\rm d}^{6.4} 
m_{\rm BH}^{ \frac{2}{3}} \le 1
 \end{equation}

Therefore the maximum donor mass that leads to transient behavior in BH-WD binaries is:
 \begin{equation}
m_{\rm tr} = 0.038\ m_{\rm BH} ^{- 0.1} 
\label{mtr}
 \end{equation}

The time interval in Gyr needed for the WD donor mass 
to evolve from $m_{d}^{-11/3}(T_1)$ to $m_{d}^{-11/3}(T_2)$ is (using eq.~\ref{dm_wd}):

 \begin{eqnarray}
 T_2 - T_1 & = & \frac{3}{22} \times 10^{-6}  m_{\rm BH}^{-2/3} \times \nonumber \\
 & & \left ( m_{d}(T_2)^{-11/3}-m_{d}(T_1)^{-11/3} \right )
\label{time_int}
 \end{eqnarray}
 Here we assume that the mass of the BH is constant, since $m_{\rm BH}\gg m_{d}$.
 Consequently and using eq.~(\ref{mtr}) we can find that the time a BH-WD system 
spends in the persistent state is 
$t_{\rm pers}\simeq\,20\,\times\,10^6 \, m_{\rm BH}^{-0.3}$ \,yr, i.e., 
it very weakly depends on the accretor mass (the dependence on the initial 
donor mass is negligible, below that 1\%).
We note that through this persistent phase the MT rate will be comparable 
or higher to the Eddington limit only for a very short time, 
$t_{\rm Edd}\simeq \,2\,\times\,10^6 \, m_{\rm BH}^{-0.9}$ \,yr 
\footnote{Although MT is non-conservative during the super-Eddington
accretion, and eq.~(\ref{jcons}) formally should not be applied, this result is 
well consistent with the detailed calculations that take into account
non-conservative MT}.

The evolution of BH-WD systems with C/O WD companions is rather similar. The critical MT rate (using Menou et al. 2002) is
 \begin{equation}
\dot m_{\rm crit} = -9.4 \times 10^{-13} m_{\rm d}^{-1.47} \ M_\odot {\rm yr^{-1}} 
\label{mcrit_co}
\end{equation}
and
 \begin{equation}
m_{\rm tr} = 0.03\ m_{\rm BH} ^{- 0.1} \ .
\label{mtr_co}
\end{equation}
 The time that a BH-WD system with a C/O rich donor spends  in the persistent state is also not very long,
$t_{\rm pers}\simeq\,50\,\times\,10^6 \, m_{\rm BH}^{-0.3}$ \,yr. 

We  conclude  that  BH-WD  binaries  that contribute  to  the  current
upper-end   XLFs  of   ellipticals  are   expected  to   be  transient
sources\footnote{This would not be true if BH-WD binaries continuously
formed, but this is not  possible in the galactic field of ellipticals
and  is not  even expected  in globular  clusters, since  BHs  tend to
dynamically  separate from  the  rest  of the  cluster  and eject  one
another
\citep{1993Natur.364..421K,1993Natur.364..423S,2000ApJ...539..331W}.}

\section{Mass Spectrum Weighting Factor and Transient Duty Cycle}

In  the previous  section  we have  shown  that the  upper-end XLF  of
ellipticals is dominated by transient  BH XRBs possibly with a variety
of donors: MS  (for the case of the IT MB  prescription) and RG stars,
and WD  donors with masses lower than  $\sim 0.035$\,M$_{\odot}$.  All
these systems contribute to the  XLF only when in outburst, when their
$L_{X}\simeq L_{\rm Edd}\propto M_{\rm BH}$. Consequently the slope of
the upper-end XLF can serve as a footprint of the BH mass distribution
of accretors in  the contributing BH XRBs. These  contributing BH XRBs
are just  a sub-set (those in  outburst) of the true  population of BH
XRBs  in ellipticals  determined by  the duty  cycle of  BH transients
binaries.  For the  general case  of a  transient duty  cycle  that is
dependent on the BH accretor mass (and possibly other quantities), the
differential  XLF   $n(L)_{\rm  obs}$  and  the   underlying  BH  mass
distribution in XRBs $n(m)_{\rm BH}$ are connected by:
 \begin{equation}
n(L_{X})_{\rm obs} = n(m_{\rm BH}) \times W(m_{\rm BH}), 
\label{nor}
 \end{equation}
 where $W(m_{\rm BH})$ is a weighting factor related to the dependence of the transient duty cycle on $m_{\rm BH}$. 
 
The observed  slope of the differential upper-end  XLF is $\alpha_{\rm
d}=2.8\pm0.6$:  $n(L_{X})_{\rm  obs}\propto\,L_{X}^{-\alpha_{\rm  d}}$
(the slope of the cumulative upper-end XLF reported by Kim \& Fabbiano
2004   is  $\alpha_{\rm   c}=1.8\pm0.6$).   Assuming  that   $n(m_{\rm
BH})\propto\,m_{\rm  BH}^{-\beta}$  and $W(m_{\rm  BH})\propto\,m_{\rm
BH}^{-\gamma}$,  the  slope  characterizing  the  underlying  BH  mass
distribution in XRBs is:
 \begin{equation}
 \beta = \alpha_{\rm d} - \gamma.
 \end{equation}
  For   the   standard   assumption   of  a   constant   duty   cycle,
$\beta=\alpha_{\rm  d}= 2.8\pm0.6$.  In  the following  subsections we
derive $W(m_{\rm BH})$  and $\gamma$ for all three  types of donors in
one example case  of a duty cycle dependent on  the binary and MT
properties (see, e.g., eq.\ref{eq:eta}).

\subsection{Main Sequence Donors} 

In  what  follows  we  estimate  the  typical  duty  cycle  for  BH-MS
transients averaged  over the  possible distribution of  donor masses.
In the case of the IT  MB perscription, the angular momentum loss rate
due to  GR is  comparable or even  more important  than MB for  all BH
masses above 3  $M_\odot$ and donor masses $\la  1.0-1.2 M_\odot$. The
MT timescale  during the  transient phase is  longer than  the donor's
thermal timescale  when the donor is $\ga  0.25$\,M$_\odot$.  For this
range the  donor is in  thermal equilibrium and the  approximation for
the  mass-radius  dependence eq.~(\ref{mr-rel})  can  be used.   Using
eq.~(\ref{jcons}),  (\ref{jgr}) and  (\ref{mcrit_h}), and  the fitting
formula for the Roche lobe radius from \citet{1983ApJ...268..368E}, we
find:
 \begin{equation}
\frac{\dot m}{\dot m_{\rm crit}} \simeq 0.054 
{ \frac { m_{\rm BH}^{0.4} (q_{\rm BH}^{-2/3}\log(1+q_{\rm BH}^{1/3}) + 0.6)^{2.1}} 
{m_{\rm d}^{2} (1+q_{\rm BH}) (4/3-1/q_{\rm BH})}} \, ,
\label{dm_dmcrit_bhms}
 \end{equation}
where $q_{\rm BH}=m_{\rm BH}/m_{\rm d}$ is the mass ratio.
For $q_{\rm BH} \gg 1$ we have

\begin{equation}
\frac{\dot m}{\dot m_{\rm crit}} \simeq 0.014 {m_{\rm BH}^{0.4}} m_{\rm d}^{-2}
\label{dm_dmcrit_bhms_short}
 \end{equation}

At donor masses smaller than $0.25$\,M$_\odot$, the donor is out of the thermal equlibirum and its radius is about twice bigger
than predicted by eq.~(\ref{mr-rel}). We then find: 
 \begin{equation}
\frac{\dot m}{\dot m_{\rm crit}} \simeq 0.0004  {m_{\rm BH}^{0.4} }
 m_{\rm d}^{-2} 
\label{dm_dmcrit_bhms_small_mass}
 \end{equation}
We note that the split into the two expressions above is a rough, but useful approximation.
 
 As  discussed in  \S\,2.2, for  IT MB,  a BH-MS  system  is transient
throughout the  MT phase.  The MT rates  are well below  the Eddington
limit  for the  BH mass  and  therefore we  assumed that  MT is  fully
conservative; i.e.,  $M_{\rm d}+M_{\rm BH} = M_{\rm  tot}$ is constant
with  time.  In what  follows  we assume  a  {\it  flat} current  mass
distribution for donors  (${\partial N}/ {\partial m_{\rm d}}=const$).
We  are  guided in  this  choice  by  results from  binary  population
synthesis  calculations  (with   the  StarTrack  code;  Belczynski  et
al.  2002  and  2005;  Belczynski  2005,  private  communication).  We
integrate   eq.~\ref{dm_dmcrit_bhms_short}   for   $m_{\rm  d}$   from
$\simeq\,0.25\, {\rm to}\, 1$\,M$_\odot$ and using eq.~(\ref{eq:eta}),
we  find that  at present  the  probability that  a system  with a  BH
accretor of $m_{\rm  BH}$ is in outburst and  therefore contributes to
the upper-end XLF is:
 \begin{equation}
W(m_{\rm BH}) = {\frac {\int_{0.25}^{m_{\rm TO}}  \eta {\frac{\partial N} {\partial m_{\rm d}}} \, d m_{\rm d}} 
{\int_{0.25}^{m_{\rm TO}} {\frac{\partial N}  {\partial m_{\rm d}}}  \, d m_{\rm d}}} \simeq 0.05 m_{\rm BH}^{0.4}, 
\label{w_for_bhms}
 \end{equation}
 where $m_{\rm TO}$ is the turn-off 
MS mass for the elliptical galaxy in solar units. 
 We note that this result is valid only for large mass ratios $q_{\rm BH}>>1$.
The contribution of BH-MS system when donors have masses $\la 0.25 M_\odot$ 
(systems where the donor is out of the thermal equilibrium) 
is less significant.

It is also important to note here that for MS donors the factor $W$ 
does not appear to depend on time (i.e., the age of the elliptical 
galaxy). 
Such a time dependence would enter in relation to the value of the maximum 
donor mass (turn-off mass for the host galaxy). 
However we find that $W(m_{\rm BH})$ is a very weak function 
of $m_{\rm TO}$, and therefore it is not sensitive to the elliptical age.

\subsubsection{Monte Carlo Simulations} 

\label{monte-carlo}

In principle, prolonged  mass accretion onto the BHs  can affect their
mass spectrum.  Since this effect cannot be  included analytically, we
have examined it quantitatively  using simple Monte Carlo simulations.
We set up the simulations assuming  a flat BH-MS birth (MT onset) rate
and a flat mass distribution for  donors at the onset of the MT phase,
without any restrictions on the mass ratio $q_{\rm BH}$.  Donor masses
at birth  were varied  in from 0.1\,$M_\odot$  to the  current $m_{\rm
TO}$  at  an  elliptical  age  of $10$\,Gyr,  assumed  as  a  standard
value. Each  BH-MS binary  was evolved to  the current  elliptical age
using equations  shown in the \S\,2.2.   For the MT  evolution we took
into  account both IT  MB and  GR, and  for the  Roche lobe  radius we
adopted the approximation by Eggleton (1983). A binary is removed from
the MT population if the  donor mass falls below $0.05$\,M$_\odot$. If
the MT  timescale is longer than  the thermal timescale  of the donor,
the donor radius evolution is simply proportional to the mass lost due
to MT.  On the  other hand, if  the MT  timescale is shorter  than the
donor's  thermal   timescale,  the  donor   is  out  of   the  thermal
equilibrium. In this case we  modify the evolution of the donor radius
using a prescription that is  in acceptable agreement with our results
from detailed  MT calculations with  the stellar evolution:  $\delta r
\sim \delta m \sqrt{\dot m_{\rm TH} / \dot m}$, where $\dot m_{\rm TH}
= m_{\rm d}/ t_{\rm TH}$ is  the MT rate driven on the donor's thermal
timescale $t_{\rm TH}$. Evolution of the transient systems follows the
adopted duty cycle (eq.[\ref{eq:eta}] with $\delta=1$).

\begin{figure}
{\plotone{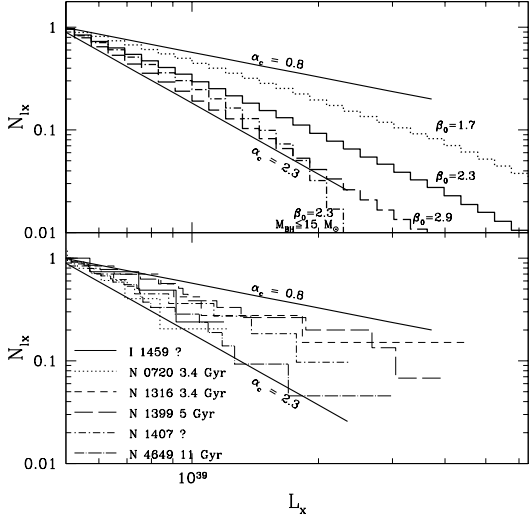}}
\caption{ The upper panel shows  model XLFs for BH-MS binaries.  Lines
shows results  for initial $\beta_0=1.7,2.3,2.9$  at an age of  10 Gyr
(dotted, solid  and dashed lines; respectively).  The dash-dotted line
is for a  $15 \,M_{\odot}$ BH mass cut-off  with $\beta_0=2.3$.  Thick
solid   lines   corresponds  to   slopes   $\alpha_{\rm  c}=0.8$   and
$\alpha_{\rm c}=2.3$.   The lower panel  shows XLFs in  observed early
type galaxies, where it is shown  that the observed range of slopes is
also within the range $\alpha_{\rm c}=0.8 - 2.3$.  Data are taken from
\cite{2004ApJ...611..846K} and  the ages  of the ellipticals  are from
\cite{2001MNRAS.326.1141R} and \cite{2005ApJ...622..235T}.  }
\label{xlf_art}
\end{figure}

Based on the results of our  Monte Carlo simulations we find that: (i)
due to accretion the BH  mass spectrum slope increases by about $0.2$,
i.e., $\beta  = \beta_{\rm  0}+0.2$, where $\beta_{\rm  0}$ is  the BH
mass slope at MT onset; (ii) the  slope of the BH mass spectrum at the
beginning  of  mass transfer  best  reproduces  the observations  with
$\beta_{0}=2.3\pm0.6$ (see Fig.~\ref{xlf_art}).  We also find that the
relation between $\beta$  and $\beta_{0}$ is not sensitive  to the age
of the elliptical (as  long as it is in the range  of a few to several
Gyr).

Next  we examine  how  the results  in  this section  are affected  by
plausible   variations  of  a   number  of   possibly  oversimplifying
assumptions  made so  far. Although  it is  possible to  re-derive the
analytical  expressions   with  different  assumptions,   it  requires
repeating  essentially  the same  analysis  multiple times,  something
inappropriate for  presentation here. Instead it is  more efficient to
examine  these  effects using  the  Monte  Carlo  simulations for  XLF
slopes.

For the tests that follow we adopt $\beta=2.3$ and examine how the 
cumulative XLF slopes $\alpha_c$ are affected.

\begin{figure}
{\plotone{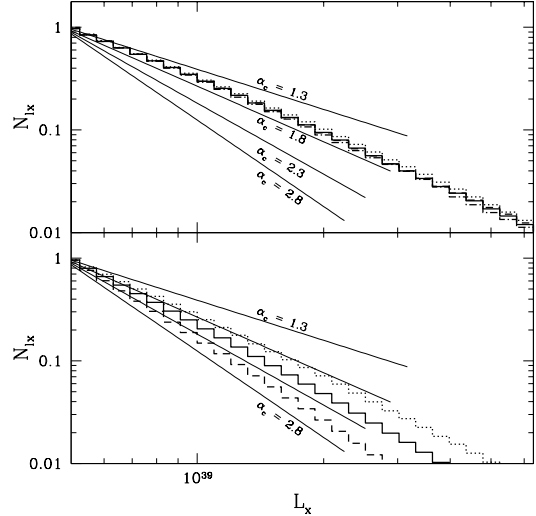}}
\caption{ Model  XLFs for BH-MS binaries for  initial $\beta_0=2.3$ at
an age of 10 Gyr. The upper panel includes results for $\eta$ given by
eq.\,(\ref{eq:eta})  for  $\delta=1$   and  outburst  peak  luminosity
deviating  by a  factor of  2  from the  Eddington luminosities  (the dash-dotted
line),  and for $\delta=0.5,  1, 2$  without any  luminosity variation
(dotted,  solid, and dashed  lines, respectively).  The bottom  panel includes
results for  $\eta$ given by eq.\,(\ref{eq:eta2})  for $\delta=0.5, 1,
2$  without  any  luminosity  variation  (dotted,  solid, and dashed  lines,
respectively). Straight lines with  various $\alpha_c$ slopes are also
shown.}
\label{xlf_art2}
\end{figure}

We first examine  the effect of random variations by a  factor of 2 of
outburst peak X-ray luminosities  of individual transient systems. The
results are shown in Fig.\,4 (top panel): the solid line is our standard
model where  peak $L_X$ are  set equal to  $L_{\rm Edd}$ (same  as the
solid curve in  Fig.~\ref{xlf_art} top panel) and the dash-dotted line is the
result we obtain when random $L_X$ variations are introduced. They are
essentially indistinguishable,  and therefore  the derivation of  a BH
mass  spectrum  slope  from  the  observed XLF  slope  appears  highly
robust. This is consistent  with the findings of observational studies
of such variations (Zezas et al. 2004).

Next we examine the effect of $\delta$ values in the case of the MT-dependent duty cycle 
in eq.\,\ref{eq:eta} other than 0 or 1. A set of curves for $\delta=$0.5, 1, and 2 
are shown in Fig.\,4 (top panel). Once again we conclude that the variations are 
essentially negligible to such quantitative changes. Instead the slope behavior 
seems to be dominated by the qualitative character of $\eta$ in eq.\,\ref{eq:eta}: 
the stronger the transient character, the smaller the duty cycle. 

Last we  examine the effect  of changing the functional  dependence of
the duty cycle normalizing it to  the Eddington MT rate instead of the
critical  rate  for  transient  behavior  (eq.\,[\ref{eq:eta2}]).  The
results are shown in Fig.\,4 (bottom panel) for $\delta=$0.5, 1, and 2
and  they  are compared  to  the our  standard  case  of $\eta$  (from
eq.\,[\ref{eq:eta}]  with $\delta=1$).  This  is the  only case  where
significant variations  are evident. More  specifically, the resultant
XLFs are steeper (absolute $\alpha_c$  values are higher) and they are
more dependent on  the choice of $\delta$. Given  that the results are
shown  for  a fixed  $\beta$  value, it  means  that,  for a  specific
observed XLF slope, this different form of the MT-dependent duty cycle
would lead to the derivation of a flatter BH mass spectrum compared to
that  derived  using eq.\,(\ref{eq:eta2}).   We  conclude that  better
understanding of the transient duty cycle and its dependence on binary
or  MT properties  is  important for  obtaining reliable  quantitative
results in the future.

\subsection{Red Giant Donors} 

In principle we should repeat the above estimate for BH-RG transients. 
However, from eq.~(\ref{mcrit_h}), 
(\ref{dmrg}) and (\ref{rdisk}) we obtain:
 \begin{equation}
\frac{\dot m_{\rm RG}}{\dot m_{\rm crit}} \simeq 0.2 a^{-0.7} m_{\rm d}^{1.5}, 
\label{dm_dmcr_rg}
 \end{equation}
 and it  is evident  that, in this  case, the MT-dependent  duty cycle
$\eta$   (see    eq.~\ref{eq:eta})   is   independent    of   the   BH
mass.   Therefore,  for   RG  donors,   $\gamma=0$.   Instead   it  is
significantly dependent  on the RG  donor mass.  Since  in ellipticals
the typical mass  for RG donors is  about the same as the  mass of the
turn-off of  MS stars,  the duty  cycle for RG  donors depends  on the
turn-off mass, and hence on the  age T [Gyr] of the elliptical.  For a
solar   metallicity   and   stars   of   $1$\,M$_\odot   \la   M   \la
1.5$\,M$_\odot$, the approximate dependence  of the turn-off mass with
the age is $m_{\rm d}\sim  m_{\rm TO}\sim 2\, T^{-1/3}$ \citep[here we
used         the          simplified         evolutionary         code
from][]{2000MNRAS.315..543H}.     Using     eq.~(\ref{eq:eta}),    and
integrating eq.~(\ref{dm_dmcr_rg}) over binary separations (similar to
our integrations for MS donors; see eq.~\ref{w_for_bhms}), we find:
 \begin{equation}
W(T) \simeq 0.03  \, T^{-0.5}, 
\label{w_for_rg}
 \end{equation}
Here  we  assume  a  distribution  of orbital  separations  for  BH-RG
binaries before  MT starts that is  flat in the  logarithm. The reason
for this  choice is that this  is appropriate for  the distribution of
zero-age  binaries and the  shape is  actually preserved  through wind
mass loss,  common-envelope evolution, and  asymmetric explosions with
small kicks appropriate for black holes (Kalogera \& Webbink 1998).

It is interesting to note that for RG donors $W$ is dependent on the 
galaxy age whereas for MS donors the dependence on the BH mass dominates.

\subsection{White Dwarf Donors} 
\label{sec_w_for_wd}

During the transient stage, the WD mass can be written as a function 
of time $T$ in Gyr (using eq. \ref{time_int}):
 \begin{eqnarray}
m_{\rm d}(T)= 0.0134  T^{-3/11} m_{\rm BH}^{-2/11} 
\label{mdage}
 \end{eqnarray}
 We  can  then  calculate  the  probability  for  a  BH-WD  system  to
contribute to  the upper-end XLF at  an elliptical age  $T$. We assume
that  (i) all  accreting  BH-WD  systems were  formed  within a  short
interval of elliptical ages $T_{\rm start}$ to $T_{\rm fin}$ (in Gyrs)
several  Gyrs ago  when  star  formation was  still  occurring in  the
elliptical; and (ii) $T-T_{\rm  fin}>t_{\rm pers}$,i.e., the binary is
a transient at time $T$. The latter assumption is well justified given
the short duration of the  persistent phase (see \S\,2.4).  We further
adopt  a constant  BH-WD formation  rate between  $T_{\rm  start}$ and
$T_{\rm  fin}$, i.e.,  ${\frac{\partial N}  {\partial  t}}=const$. The
probability then  is expressed by  the duty-cycle weighting  factor at
$T$ for a given BH mass:
 \begin{eqnarray}
W(T; m_{\rm BH}) & = & {\frac {\int_{m_{\rm d1}}^{m_{\rm d2}}  \eta {\frac{\partial N} {\partial m_{\rm d}}} \, d m_{\rm d}} 
{\int_{m_{\rm d1}}^{m_{\rm d2}} {\frac{\partial N}  {\partial m_{\rm d}}}  \, d m_{\rm d}}}  \nonumber \\
& = & 0.1 {\frac 
{\int_{m_{\rm d1}}^{m_{\rm d2}} 
{\frac{\dot m_{\rm d}}{\dot m_{\rm crit}}} 
\frac{\partial N} {\partial t} 
\frac{\partial t} {\partial m_{\rm d}}  \, d m_{\rm d}}
{\int_{m_{\rm d1}}^{m_{\rm d2}} {\frac{\partial N}  {\partial t}} {\frac{\partial t}  {\partial m_{\rm d}}} \, d m_{\rm d}}} \nonumber \\
& = & 0.1 {\frac {\int_{m_{\rm d1}}^{m_{\rm d2}} \dot m_{\rm crit}^{-1}  \, d m_{\rm d}} {\int_{m_{\rm d1}}^{m_{\rm d2}} \dot m_{\rm d}^{-1}  \, d m_{\rm d}}}
 \end{eqnarray}
 Here $m_{\rm d1}=m_{\rm d}(T-T_{\rm start};m_{\rm BH})=m_{\rm d}(t_1;m_{\rm BH})$ and $m_{\rm d2}=m_{\rm d}(T-T_{\rm fin};m_{\rm BH})=m_{\rm d}(t_{2},m_{\rm BH})$; 
$m_{\rm d2}> m_{\rm d1}$.
Then, using eq.~(\ref{dm_wd}), (\ref{mcrit_he}) and (\ref{mdage}),  we obtain:
 \begin{eqnarray}
W(T;m_{\rm BH}) & = & 1.6 \times 10^{-4} m_{\rm BH}^{-0.5} t_1^{-7/4} \nonumber \\
& & \times \frac{1 - (t_2/t_1)^{-3/4}} { 1 - (t_2/t_1)}
\label{w_for_bhwd}
 \end{eqnarray}
 Therefore for WD donors $\gamma=0.5$ and $W$ depends on both the BH mass and the galaxy age.

\section{Accreting Black Hole Mass Spectrum}

So  far we  have derived  the dependence  of the  duty-cycle weighting
factor $W$ (eq.\ref{nor}) on the accreting BH mass and the age of
the host elliptical galaxy, for  the different types of BH donors. In
order  to  make  progress  and  develop  a  method  for  deriving
constraints on the slope $\beta$  of the underlying accreting BH mass
spectrum we  need to examine  which of the possible  donor populations
dominate the observed upper-end  XLF under what conditions. The answer
to this question requires large-scale population synthesis models that
are not part  of the scope of this paper. However,  as is shown below,
we can use  a number of different arguments and  pieces of evidence to
derive tentative constraints. The primary purpose of our analysis
is not to derive an  unambiguous constraint at present, but instead to
develop a methodology for how to derive the most reliable constraints,
given   the  current   uncertainties  associated   with   our  current
understanding of these X-ray binaries.

For RG  donors we find that in  the case of a  MT-dependent duty cycle
$\eta$ is about  an order of magnitude smaller than  for MS donors: by
comparing  eq.~(\ref{w_for_bhms}) and  eq.~(\ref{w_for_rg}),  using an
elliptical-galaxy age of at least  a few Gyr (more typical is 12\,Gyr)
and  a BH  mass of  at least  $3$\,M$_\odot$ the  difference  with the
$\eta$ value for MS donors is a factor of 15.  We also note that among
known BH X-ray transients in  our Galaxy, the ratio of transient BH-RG
systems to  transient BH-MS systems is about  1:2 \citep[see Table~4.1
in][]{BH_book_ch4}.   Consequently we  conclude that  BH-RG transients
cannot be  important contributors to  the upper-end XLF  of elliptical
galaxies  for the  example  case  of the  MT-dependent  duty cycle  as
defined  by  $\eta$.   {\it  Only}  in  the  case   of  the  constant,
MT-independent  (and  therefore donor  and  BH-mass independent)  duty
cycle we expect BH-RG transients to be a significant population of the
observed upper-end XLF.

Let  us consider  the case  when  the number  of transient  BH-RG
systems exceeds the number of transient BH-MS system in a way that the
contributions of the two populations  become comparable at some age of
the elliptical.  We also assume that $\beta_0$ should be the same for
both populations.  In this  case, the resultant  combined XLF  will be
flatter than  the XLF provided by only  BH-MS contributors.  Secondly,
as the contribution of BH-RG system decreases with elliptical age (see
eq.~\ref{w_for_rg}),  the XLF  becomes steeper,  evolving  towards the
slope characteristic for BH-MS binaries.   It is possible that this is
the  kind of  behavior that  we observe  in XLFs  of  ellipticals (see
Fig.~3): we note that younger ellipticals appear to have flatter XLFs,
although uncertainties are significant.

For    WD    donors   we    find    $W(m)\propto   t_1^{-7/4}$    (see
eq.~\ref{w_for_bhwd}),  implying that  the probability  of  each BH-WD
transient contributing to  the observed XLF decreases with  the age of
the  elliptical galaxy  (similar to  the case  of RG  donors, but
unlike the case  of MS donors).  Let us  consider an elliptical where
the formation  of BH-WD systems has ended  at least a few  Gyr ago. We
also  consider  that  BH-MS  systems  are transient  and  have  $W(M)$
according  eq.~(\ref{w_for_bhms}).  In  order  for BH-WD  binaries  to
contribute significantly to the observed  XLF they must form at a rate
such that more than $\sim 8,000$ BH-WDs form for each BH-MS. This
ratio is calculated adopting a value  for the age of the elliptical of
$\simeq 3.5$\,Gyr  (among the lowest  reported in the  literature) and
for a choice of  BH masses in binaries with WD and  MS donors, so that
their Eddington X-ray luminosities  are comparable, and therefore they
contribute to the  same X-ray luminosity bin (3\,M$_\odot$  for WD and
5\,M$_\odot$ for MS donors).  For more typical, older ellipticals with
ages closer  to 10\,Gyr  the required ratio  becomes even  higher than
$8,000$.  According to  binary  population synthesis  models for  the
Milky Way published  so far, the number of  formed BH-WD LMXBs exceeds
the   number  of   BH-MS   LMXBs  by   at   most  a   factor  of   100
\citep{2002MNRAS.329..897H}.  Furthermore, the lifetime  of BH-MS
binaries is of  order 1\,Gyr (or a few Gyr;  see also Fig.~1), whereas
the lifetime  BH-WD binaries is longer,  but cannot exceed  the age of
the elliptical galaxy ($\sim10$\,Gyr).  We conclude that the number of
BH-WD LMXBs  could be at most about  a factor of 1000  higher than the
number of BH-MS LMXBs, but this  ratio is still below what is required
for BH-WD  to become an important  contributor.  So, if  the ratio of
BH-MS binaries to BH-RG binaries  in ellipticals is similar to that in
the Milky Way, BH-WD XRBs will not be a significant contributor to the
XLF.  Based  again on the discrepancy  between the duty  cycles for WD
and RG donors (smaller for WDs by a factor of $\sim 800$), expect that
BH-RG transients dominate over BH-WD transients too.

For  the case  of an  example  MT-dependent duty  cycle (expressed  by
$\eta$ in eq.~\ref{eq:eta})  we conclude that: (i) if  IT MB describes
the   angular  momentum   loss  best,   then  only   BH-MS  transients
significantly   contribute  to  the   XLFs  of   elliptical  galaxies;
consequently  $\beta=2.5\pm0.6$; (ii) if  instead RVJ  MB is  a better
prescription,  then  the  XLF  is  dominated  by  BH-RG  binaries  and
$\beta=2.8\pm0.6$.

For the case  of a constant duty cycle independent  of the donor type,
it is clear  that the XRB type with the  highest formation rate should
dominate  the  XLF.   According   to  formation  rates  calculated  by
\citet{2002MNRAS.329..897H}, BH-WD  binaries form at a  rate about 100
times higher than BH-MS and BH-RG binaries. Consequently, WDs would be
expected to  dominate the transient  population and this  is certainly
not true for the Milky  Way. Therefore we conclude that the assumption
of  a  constant,  MT-independent  duty  cycle  is  most  probably  not
realistic.

Overall,  we conclude  that MS  or  RG donors  dominate, depending  on
whether the IT or RVJ MB prescription is more realistic. Consequently,
the  slope of  the  accreting BH  mass  spectrum is  $\beta=2.5\pm0.6$
($\beta_0=2.3\pm0.6$)  or $\beta=2.8\pm0.6$,  respectively. These
quantitative results  are of course  dependent on the  adopted example
form of the MT-dependent duty cycle (eq.\,[\ref{eq:eta}]). As shown in
\S\,3.1.1 a possible different form (e.g., eq.\,[\ref{eq:eta2}]) could
lead to somewhat flatter values for $\beta$. 

Next we consider the fact that the upper-end XLF of ellipticals is not  
a perfect power-law up to arbitrarily high $L_{X}$ values; 
instead there is a usually smooth cut-off behavior that limits 
the maximum $L_{X}$ observed at $\simeq 2\times10^{39}$\,erg\,s$^{-1}$. 
In Fig.~\ref{xlf_art} we show the {\em cumulative} XLF associated 
with a model population of BH-MS binaries with a BH mass spectrum 
with a {\em differential} slope of $\beta=2.3$  and  
with an imposed upper limit of $15$\,M$_{\odot}$ 
on the maximum BH mass present in the XRB population.
We obtain a model XLF that behaves very similarly to observed XLFs (dash-dotted line). 
Clearly this is just to show the importance of the qualitative effect of a BH mass cut-off on the cumulative XLF.

\section{DISCUSSION}

We consider the upper-end XLF of ellipticals (above the reported break
at  $\simeq 4-6\times10^{38}$\,erg\,s$^{-1}$) and  suggest that  it is
populated by BH X-ray transients at outburst emitting approximately at
the  Eddington limit.   We argue  that the  upper-end XLF  slope  is a
footprint of the  underlying accreting BH mass spectrum  modified by a
weighting function related to the  transient duty cycle.  We show that
this  weighting   factor  is  generally  dependent  on   the  BH  mass
and/or  the age  of the  host galaxy  and the  derived power-law
dependence is different  for each of the possible  BH donor types: MS,
RG,  and  WD. Our  predicted  dominance  of  X-ray transients  at
outburst contributing to the upper-end XLF could possibly be tested by
future   high-resolution  X-ray   observations  designed   to  achieve
long-term   monitoring   probably  at   time   scales   of  years   or
longer. Unfortunately,  given the uncertainties  in the theory  of the
thermal disk instability,  it is not possible to  make any predictions
about the expected duration of these outbursts. 

Based  on  our analysis  and  prior  population  synthesis results  we
conclude that a constant transient duty cycle independent of the donor
type  can be  excluded. Instead  a  duty cycle  dependent on  the
binary and MT properties seems to be required. Given the uncertainties
associated with transient duty cycles at present, we adopt a couple of
different       formulations      of      such       a      dependence
(eqs.\,[\ref{eq:eta},\ref{eq:eta2}]), as reasonable examples, which in
no way exhaust the possibilities. In the specific case of a duty cycle
that depends  on the  ratio of  the binary mass  transfer rate  to the
critical  rate  for   transient  behavior  (see  eq.~\ref{eq:eta})  we
conclude find that  the BH X-ray transients forming  the upper-end XLF
in ellipticals  have a  dominant donor type  and an accreting  BH mass
spectrum  slope $\beta$  that depend  on  the strength  of MB  angular
momentum loss: (i) for the  IT MB prescription, only BH-MS transients
significantly  contribute to the  upper-end XLF  and $\beta=2.5\pm0.6$
($\beta_0=2.3\pm0.6$ ); (ii)  for the RVJ MB prescription,  the XLF is
dominated by BH-RG binaries and $\beta=2.8\pm0.6$.  We note that these
quantitative  results  do  depend   on  our  conclusions  about  which
donor-type   population  dominates   based   on  currently   published
population   synthesis  models   \citep{2002MNRAS.329..897H}   and  on
available observations  of BH X-ray  systems in our Galaxy.  If, e.g.,
the  relative fraction of  BH-RG transients  in ellipticals  is larger
than  the observed  relative fraction  in our  Galaxy, we  expect that
BH-RG  binaries contribution  will lead  to a  time-dependence  of XLF
slopes,  where younger  ellipticals will  have a  slope  predicted for
BH-RG  binaries, and older  ellipticals a  steeper slop  predicted for
BH-MS binaries.

The primary  goals of this study  are to present (i)  the line of
arguments that  connects the upper-end  XLF of ellipticals to  BH XRBs
formed  in the  galactic field  and (ii)  the methodology  for  how to
extract  information about  the accreting  BH mass  spectrum  from the
observed XLF slopes. We  have further obtained quantitative results on
the BH mass spectrum  slope under certain reasonable assumptions, some
of which  (e.g., the functional  form of the MT-dependent  duty cycle)
represent mere  examples. A careful  examination of the  robustness of
these quantitative results has been for  the case of MS donors. It has
been  found that  the derived  slopes  are robust  against (i)  random
variations  by factors of  a few  of the  outburst peak  luminosity of
individual  sources,  and  against  (ii) variations  of  the  possible
duty-cycle   dependence  on   the  critical   MT  rate   for  outburst
behavior.  However,   completely  difference  duty-cycle  dependencies
cannot be excluded.  An improved understanding of this  issue would be
required to  derive reliable quantitative conclusions  about the value
of the BH mass spectrum slope in transient XRBs in ellipticals.  

We expect  that our analysis  and methods can  be used to  reveal more
information  about the formation  of BH  XRBs in  elliptical galaxies.
More  specifically they  could  eventually be  used  to constrain  the
physical connection  between massive stars in XRB  progenitors and the
resultant   BH   masses.   Current   simulations   assume  either   an
artificially  constant   mass  for   BHs  formed  (usually   at  $\sim
10$\,M$_{\odot}$),  or  a constant  mass  fraction  of the  progenitor
leading to the remnant objects,  or a remnant mass relation consistent
with core-collapse  simulations. Constraints on the  accreting BH mass
spectrum as those discussed here could contribute to our understanding
of core collapse, and the  connection of BH masses to their progenitor
masses.

\acknowledgments  We   thank  K.~Belczynski,  J.~McClintock,  and
R.~Remillard  for useful  discussions  and the  anonymous referee  for
suggestions that  greatly improved the manuscript and  motivated us to
perform a number of tests. This work is partially supported by a NASA
{\em Chandra}  Theory Award  to N.\ Ivanova  and a  Packard Foundation
Fellowship  in  Science  and  Engineering  to V.\  Kalogera.  VK  also
acknowledges  the hospitality of  the Aspen  Center for  Physics where
part of this work was completed.


\newpage

\begin{thebibliography}{}

\bibitem[Belczynski et al.(2002)]{2002ApJ...572..407B} Belczynski, K., 
Kalogera, V., \& Bulik, T.\ 2002, \apj, 572, 407 
\bibitem[Belczynski et al.(2005)]{Belcz_2005} Belczynski, K., 
Kalogera, V., Rasio, F., Taam, R., \& Ivanova 2005, in prep 
\bibitem[Bildsten \& Deloye(2004)]{2004ApJ...607L.119B} Bildsten, L., \& 
Deloye, C.~J.\ 2004, \apjl, 607, L119 
\bibitem[Dubus et al.(1999)]{1999MNRAS.303..139D} Dubus, G., Lasota, J., 
Hameury, J., \& Charles, P.\ 1999, \mnras, 303, 139 
\bibitem[Eggleton(1983)]{1983ApJ...268..368E} Eggleton, P.~P.\ 1983, \apj, 
268, 368 
 \bibitem[Fabbiano et al.\ (2001)]{FAB01} Fabbiano, A., Zezas, A., \& Murray, S. S.  2001, \apj, 554, 1035
\bibitem[Hurley et al.(2000)]{2000MNRAS.315..543H} Hurley, J.~R., Pols, 
O.~R., \& Tout, C.~A.\ 2000, \mnras, 315, 543 
\bibitem[Hurley et al.(2002)]{2002MNRAS.329..897H} Hurley, J.~R., Tout, 
C.~A., \& Pols, O.~R.\ 2002, \mnras, 329, 897 
 \bibitem[Iben \& Livio(1993)]{1993PASP..105.1373I} Iben, I.~J.~\& Livio,
M.\ 1993, PASP, 105, 1373
\bibitem[Ivanova \& Taam(2003)]{2003ApJ...599..516I} Ivanova, N., \& Taam, 
R.~E.\ 2003, \apj, 599, 516 
\bibitem[Ivanova \& Taam(2004)]{2004ApJ...601.1058I} Ivanova, N., \& Taam, 
R.~E.\ 2004, \apj, 601, 1058 
\bibitem[Gilfanov(2004)]{2004MNRAS.349..146G} Gilfanov, M.\ 2004, \mnras, 
349, 146 
\bibitem[Jord{\' a}n et al.(2004)]{2004ApJ...613..279J} Jord{\' a}n, A., et 
al.\ 2004, \apj, 613, 279 
\bibitem[Juett(2005)]{2005ApJ...621L..25J} Juett, A.~M.\ 2005, \apjl, 621, 
L25 
\bibitem[Kalogera \& Webbink(1998)]{1998ApJ...493..351K} Kalogera, V., \& 
Webbink, R.~F.\ 1998, \apj, 493, 351 
 \bibitem[Kalogera et al.(2004)]{2004ApJ...601L.171K} Kalogera, V., King, 
A.~R., \& Rasio, F.~A.\ 2004, \apjl, 601, L171 
\bibitem[King(2000)]{2000MNRAS.312L..39K} King, A.~R.\ 2000, \mnras, 312, 
L39 
\bibitem[King(2005)]{BH_book_ch13} King, A., in press,
Chapter 13 in "Compact Stellar X-ray Sources," eds. W.H.G. Lewin and M. van der Klis, Cambridge University Press.
\bibitem[King et al.(1997)]{1997ApJ...484..844K} King, A.~R., Frank, J., 
Kolb, U., \& Ritter, H.\ 1997, \apj, 484, 844 
 \bibitem[Kim \& Fabbiano(2003)]{2003ApJ...586..826K} Kim, D., \& Fabbiano, 
G.\ 2003, \apj, 586, 826 
\bibitem[Kim \& Fabbiano(2004)]{2004ApJ...611..846K} Kim, D., \& Fabbiano, 
G.\ 2004, \apj, 611, 846 
\bibitem[Kulkarni et al.(1993)]{1993Natur.364..421K} Kulkarni, S.~R., Hut, 
P., \& McMillan, S.\ 1993, \nat, 364, 421 
\bibitem[Menou et al.(2002)]{2002ApJ...564L..81M} Menou, K., Perna, R., \& 
Hernquist, L.\ 2002, \apjl, 564, L81 
\bibitem[McClintock \& Remillard(2005)]{BH_book_ch4} McClintock, J.~E.~\& Remillard, R.~A., in press,
Chapter 4 in "Compact Stellar X-ray Sources," eds. W.H.G. Lewin and M. van der Klis, Cambridge University Press.
\bibitem[Paczy{\' n}ski(1971)]{1971ARA&A...9..183P} Paczy{\' n}ski, B.\ 
1971, \araa, 9, 183 
\bibitem[Piro \& Bildsten(2002)]{2002ApJ...571L.103P} Piro, A.~L., \& 
Bildsten, L.\ 2002, \apjl, 571, L103 
\bibitem[Rappaport et al.(1983)]{1983ApJ...275..713R} Rappaport, S., 
Verbunt, F., \& Joss, P.~C.\ 1983, \apj, 275, 713 
\bibitem[Rappaport et al.(1987)]{1987ApJ...322..842R} Rappaport, S., Ma, 
C.~P., Joss, P.~C., \& Nelson, L.~A.\ 1987, \apj, 322, 842 
 \bibitem[Ritter(1999)]{1999MNRAS.309..360R} Ritter, H.\ 1999, \mnras, 309, 
360 
\bibitem[Ryden et al.(2001)]{2001MNRAS.326.1141R} Ryden, B.~S., Forbes, 
D.~A., \& Terlevich, A.~I.\ 2001, \mnras, 326, 1141 
\bibitem[Sarazin et al.(2000)]{2000ApJ...544L.101S} Sarazin, C.~L., Irwin, 
J.~A., \& Bregman, J.~N.\ 2000, \apjl, 544, L101 
\bibitem[Sarazin et al.(2003)]{2003ApJ...595..743S} Sarazin, C.~L., et al.\ 2003, \apj, 595, 743
\bibitem[Sigurdsson \& Hernquist(1993)]{1993Natur.364..423S} Sigurdsson, 
S., \& Hernquist, L.\ 1993, \nat, 364, 423 
\bibitem[Tanaka \& Shibazaki(1996)]{1996ARA&A..34..607T} Tanaka, Y., \& 
Shibazaki, N.\ 1996, \araa, 34, 607 
\bibitem[Temi et al.(2005)]{2005ApJ...622..235T} Temi, P., Mathews, W.~G., 
\& Brighenti, F.\ 2005, \apj, 622, 235 
\bibitem[Verbunt \& Zwaan(1981)]{1981A&A...100L...7V} Verbunt, F.~\& Zwaan,
C.\ 1981, A\&A, 100, L7
 \bibitem[Verbunt(1993)]{1993ARA&A..31...93V} Verbunt, F.\ 1993, \araa, 31, 
93 
 \bibitem[Watters et al.(2000)]{2000ApJ...539..331W} Watters, W.~A., Joshi, 
K.~J., \& Rasio, F.~A.\ 2000, \apj, 539, 331 
\bibitem[Webbink et al.(1983)]{1983ApJ...270..678W} Webbink, R.~F., 
Rappaport, S., \& Savonije, G.~J.\ 1983, \apj, 270, 678 
\bibitem[Weisskopf et al.(2003)]{2003ExA....16....1W} Weisskopf, M.~C., et 
al.\ 2003, Experimental Astronomy, 16, 1 
\bibitem[Zezas et al.(2004)]{2004RMxAC..20...53Z} Zezas, A., Fabbiano, G., 
Baldi, A., King, A.~R., Ponman, T.~J., Raymond, J.~C., \& Schweizer, F.\ 
2004, Revista Mexicana de Astronomia y Astrofisica Conference Series, 20, 
53 
\end{thebibliography}
\end{document}